\documentclass[aps,pra,superscriptaddress,twocolumn,floatfix]{revtex4}

\usepackage{amsmath}
\usepackage{amssymb}
\usepackage{bm,color}
\usepackage{graphicx,epsfig,color}
\usepackage{hhline}
\usepackage{ifthen}
\usepackage[utf8]{inputenc}

\definecolor{orange}{RGB}{245,31,21}

\newcommand{\bra}[1]{\langle\,{#1}\, |}
\newcommand{\ket}[1]{|\,{#1}\,\rangle}

\newcommand{\ee}{\vec{e}}
\newcommand{\kk}{\vec{k}}

\graphicspath{{pics/png/}{pics/eps/}{../2D_background_absorption_dist_4_2015_04_24/}}

\newcommand{\Deriv}[2][\empty]{%
  \ifthenelse{\equal{#1}{\empty}}
    {D_#2}
    {D_{#1,#2}}
}

\newcommand{\psit}[1][\empty]{%
  \ifthenelse{\equal{#1}{\empty}}
    {\psi_t}
    {\psi_t^{(#1)}}
}
\newcommand{\npsit}[1][\empty]{%
  \ifthenelse{\equal{#1}{\empty}}
    {\tilde\psi_t}
    {\tilde\psi_t^{(#1)}}
}
\newcommand{\rhot}[1][\empty]{%
  \ifthenelse{\equal{#1}{\empty}}
    {\rho_t}
    {\rho_t^{(#1)}}
}

\let\emph\relax 
\DeclareTextFontCommand{\emph}{\em}

\begin{document}

\title{Hierarchy of equations to calculate the linear spectra of molecular aggregates -- Time-dependent and frequency domain formulation}

\author{P.-P. Zhang}
\affiliation{Max Planck Institute for the Physics of Complex Systems,
N\"othnitzer Strasse 38, D-01187 Dresden, Germany}

\author{Z.-Z. Li}
\email{zengzhao@pks.mpg.de}
\affiliation{Max Planck Institute for the Physics of Complex Systems,
N\"othnitzer Strasse 38, D-01187 Dresden, Germany}

\author{A.\ Eisfeld}
\affiliation{Max Planck Institute for the Physics of Complex Systems,
N\"othnitzer Strasse 38, D-01187 Dresden, Germany}

\date{\today}

\begin{abstract}
In a recent publication [J.~Chem.~Phys.~142, 034115 (2015)] we have derived a hierarchy of coupled differential equations in time domain to calculate the linear optical properties of molecular aggregates.
Here we provide details about issues concerning the numerical implementation.
In addition we present the corresponding \emph{hierarchy} \it{in} \textit{frequency} domain.
\end{abstract}
\maketitle

\section{Introduction}
Molecular aggregates are assemblies of molecules where interaction between transition dipoles of different molecules (monomers) leads to a delocalization of electronic excitation over several monomers \cite{Ko96__,WueKaSa11_3376_,SaEiVa13_21_}. 
Linear optical properties (absorption, linear- and circular-dichroism) can provide information about the degree of excitonic delocalization,  the arrangement of the monomers and the strength of the dipole-dipole couplings. 
However, the interpretation of the spectra is complicated by the coupling to vibrational modes \cite{EiSeEn08_186_}.
The inclusion of vibrational modes makes the calculation of aggregate spectra quite challenging and various methods have been developed to handle vibrations (see for example~\cite{WiMo60_872_,Foe65_93_,BrHe71_865_,He77_1795_,LuFr77_36_,FrSi81_1166_,ScFi84_269_,KReMa96_99_,DaKoKl02_031919_,BeDaKj02_5810_,YaFl02_163_,Ei07_321_,FiSeEn08_12858_,IsFl09_234111_,Sp09_4267_,PrChHu10_050404_,RoStEi11_034902_,MoZhCa12_115412_,ZhHuZh15_014113_}).
In particular, the numerical treatment of many weakly damped molecular vibrations  is demanding \cite{RoEiDv11_054907_}. 
Additional complications arise, when finite temperature effects have to be accounted for. 

In Ref.~\cite{RiSuMoe15_034115_} some of us  have developed a hierarchy of differential equations that allows one to efficiently handle this situation.
The method is based on an open quantum system description, where electronic (excitonic) degrees of freedom are included in the system  and all vibrational modes are part of the environment.
From the solution of the hierarchy one can then construct the time-dependent dipole-correlation function, whose Laplace transformation is related to absorption and dispersion spectra.

While in Ref.~\cite{RiSuMoe15_034115_} the focus was on the derivation of the hierarchy at finite temperature, in the present work we discuss aspects on the numerical implementation of this hierarchy.
In addition  we show that the time-dependent hierarchy of Ref.~\cite{RiSuMoe15_034115_} can be directly transformed to frequency domain to obtain a hierarchy of coupled linear equations, instead of coupled differential equations.
This hierarchy provides a different viewpoint and offers an alternative way of numerically calculating the spectrum. 
It could also help to develop new analytical approximations. 

The paper is organized as follows:
first we briefly provide the relevant formulas used to model the aggregate.
In Section \ref{sec:Method} we present the time-dependent hierarchy of Ref.~\cite{RiSuMoe15_034115_} and its relation to the absorption spectrum in a concise manner. 
In this section we also derive the hierarchy in frequency domain and discuss some aspects of the numerical implementation of both hierarchies.
Finally we conclude by relating to the hierarchy used in Ref.~\cite{RoStEi11_034902_} for the calculation of absorption spectra.

\section{Model \label{sec:Model}}
We briefly repeat  our model of the aggregate.
Details can be found in  Ref.~\cite{RiSuMoe15_034115_}.
For each monomer we take two electronic states into account, the ground state $\ket{\phi^g_n}$ and the excited state $\ket{\phi^e_n}$; here $n$ labels the monomer.
The electronic ground state of the aggregate is simply a product state where all monomers are in their electronic ground state.
Since we are interested in linear optical properties, we will restrict the excited state basis to states 
$
 \ket{\pi_n}=\ket{\phi^e_n}\prod_{m\ne n}^N \ket{\phi^g_m},
$
where one monomer is electronically excited (e) and all other monomers are in the ground electronic state (g).
The excited state Hamiltonian is then given by
\begin{equation}
\label{H_sys}
  H_{\rm agg}=\sum_{n=1}^N\varepsilon_n\ket{\pi_n}\bra{\pi_n}+\sum_{n,m=1}^N V_{nm}\ket{\pi_n}\bra{\pi_m},
\end{equation}
where $\varepsilon_n$ is the transition energy of monomer $n$ and the matrix element $V_{nm}$ causes electronic excitation to be transferred from monomer $m$ to monomer $n$.
For later use we define the matrix-elements 
$(H_{\rm agg})_{nm}\equiv \bra{\pi_n}H_{\rm agg}\ket{\pi_m}=\varepsilon_n \delta_{nm}+V_{nm}$. 
The coupling to vibrational modes
$H_{\rm env} =\sum_{n=1}^N\sum_{\lambda}\omega_{n\lambda}a^{\dagger}_{n\lambda}a_{n\lambda} 
$
 (here  $\lambda$ denotes the modes of the monomers) 
is given by
\begin{equation}
\label{HInt}
  H_{\rm int} = \sum_{n=1}^N L_n \sum_{\lambda}\kappa_{n\lambda}(a^{\dagger}_{n\lambda}+a_{n\lambda})
\end{equation}
with the  operators $L_n = \ket{\pi_n}\bra{\pi_n}$ and $\kappa_{n\lambda}$ describes the coupling strength to mode $\lambda$ of monomer $n$.
The connection of this model to molecular vibrational modes is for example discussed in Refs.~\cite{MaKue11__,Mu95__,RoStWh12_204110_}.
It is convenient to introduce the so-called spectral density $\mathcal{J}_n(\omega)= \sum_{\lambda}|\kappa_{n\lambda}|^2 \delta(\omega-\omega_{\lambda})$. For molecules in solution the spectral density can usually be considered to be a continous function of $\omega$. Typical forms of the spectral density are discussed e.g.~in Ref.~\cite{RiEi14_094101_}.  
To treat finite temperature we introduce the  so-called `bath-correlation function' \cite{MaKue11__,Mu95__}
\begin{equation}
\label{eq:def:alpha_with_lambda}
\begin{split}
\alpha_n(\tau) = \int_0^{\infty} {\mathrm d}\omega \mathcal{J}_n(\omega) \Big( \coth\big(\frac{\omega}{2T}\big) \cos(\omega \tau)  -i\sin(\omega \tau) \Big)
\end{split}
\end{equation}
where $T$ is the temperature and we have set the Boltzman constant $k_{\rm B}=1$ (we also set $\hbar=1$ throughout the work). 
For a practical implementation we approximate the bath correlation functions $\alpha_n(\tau)$ given in Eq.~(\ref{eq:def:alpha_with_lambda}) as sums of exponentials
\begin{equation}
\label{eq:def:alpha_decomposition}
 \alpha_n(\tau) =\sum_{j=1}^{J}  p_{nj} e^{i\omega_{nj}\tau}; \quad \tau \ge 0 
\end{equation}
with complex frequencies $\omega_{nj}= \Omega_{nj}+ i \gamma_{nj}$  and  prefactors $p_{nj}$ that may also be complex.

\section{The Method}
\label{sec:Method}

\subsection{Time domain formulation}
In the following we treat the absorption spectrum as an example.
Linear and circular dichroism can be treated analogously.
 The transition strength at frequency $\omega$ can be obtained from a half-sided Fourier transformation \cite{Mu95__, MaKue11__}
\begin{equation}
\label{eq:spectrum-time}
 F(\omega) = \mathrm{Re} \; \int_0^{\infty} dt \, e^{i \omega t}\, c(t),
\end{equation}
with
\begin{align}
\label{eq:def:C_abs-time}
 c(t) =\sum_{nm} \vec{\mu}_n\cdot \vec{\mu}_m \, \mathcal{C}_{nm}(t).
\end{align}
Here $\vec{\mu}_n$ is the transition dipole moment between the ground and excited state of monomer $n$.
To obtain the functions $\mathcal{C}_{nm}(t)$  the following hierarchy is solved (here $n$ represents the monomer indices and $j$ runs over the `modes' of the bath-correlation function Eq.~(\ref{eq:def:alpha_decomposition})): 
\begin{eqnarray}
\label{eq:C(t)_HOPS}
  \partial_t\Psi_n^{(\kk)}(t) &=&
  -i \sum_m \big(  (H_{\rm agg})_{nm} - \delta_{nm} \sum_{ j} k_{n j} \omega_{n j} \big)\,\Psi_m^{(\kk)}(t) \notag \\
  && +  \sum_j k_{n j} p_{n j} \Psi_n^{(\kk-\ee_{n j})}(t)\notag \\
  && -  \sum_j \Psi_n^{(\kk+\ee_{n j})}(t),
\end{eqnarray}
where $\kk=\{k_{11},\dots,k_{NJ}\}$, with $k_{n j}$ integers $\ge 0$ and $\ee_{n j}=\{0,\dots,1,\dots 0\}$ is a vector that has a one at the ($n, j$)th position and the rest of the elements are zero.

The desired functions $\mathcal{C}_{nm}(t)$ are then obtained from
$\mathcal{C}_{nm}(t)=\Psi_n^{(\vec{0})}(t)$
where the initial state is chosen according to
\begin{equation}
\label{eq:initial_state}
\Psi_n^{(\vec{k})}(0)=\left\{
\begin{array}{lr}
\delta_{nm} &  \vec{k} = \vec{0}
\\
0, & \quad \quad \quad \mathrm{otherwise}
\end{array}
\right.
\end{equation}

\subsubsection{Comments on numerical implementation}
In numerical implementations the hierarchy has to be truncated.
Possible truncation schemes are discussed in Ref.~\cite{SuEiSt14_150403_}, see also the appendix of Ref.~\cite{RoEiDv11_054907_}. 
In simple truncation schemes one sets the last line of Eq.~(\ref{eq:C(t)_HOPS}) equal to zero if the vector $\vec{k}$ fulfills a certain condition (e.g.~$\sum_{n,j} |\omega_{nj}| k_{nj}<E_{\rm max}$, where $E_{\rm max}$ has to be chosen to be sufficiently large to ensure convergence).
The number of coupled equations (which we will denote as the size of the hierarchy) depends sensitively on the way the hierarchy is truncated.
Different truncation schemes that have the same size can result in various accuracies.
 Convergence can usually be checked by increasing the size of the hierarchy (e.g.\ by increasing $E_{\rm max}$ in the above mentioned truncation scheme). 
We want to emphasize that for a large number of molecules and for a large number of modes $J$ it is important to have a good truncation scheme.

From our experience, the hierarchy of differential equation (\ref{eq:C(t)_HOPS}) can be solved by standard propagators (like Runge-Kutta). 
Note that the form of hierarchy allows a very sparse representation. 
In addition the hierarchy can be efficiently and easily parallelized.

To obtain the desired spectrum in frequency space, the usual considerations of numerical Fourier transformations have to be taken into account.
The final time $T_{\rm max}$ should be large enough to obtain the desired resolution in frequency space ($\Delta \omega \sim 1/T_{\rm max}$).
If the correlation function $c(t)$ has not smoothly decayed to zero at the time $T_{\rm max}$ then one typically multiplies $c(t)$ by a window function (e.g.\ a Gaussian or an exponential; 
this leads to a convolution of the spectrum with a Gaussian or a Lorentzian in frequency domain, respectively). 
The width in frequency domain is inversely proportional to the width of the window function. 
The time interval $\Delta t$ at which $c(t)$ is recorded determines the maximal frequency range.
Typically this interval is much larger than the step size of the propagator. 
Finally, let us note that besides standard Fourier transformations one can also try other methods, like `harmonic inversion' or approaches based on `compressive sensing' \cite{KaLeSi10_FTuE3_}.

\subsection{Frequency domain formulation}
\newcommand{\Lap}[1]{\tilde{#1}}
The half-sided Fourier transformation in Eq.~(\ref{eq:spectrum-time}) is closely related to the Laplace transformation
\begin{eqnarray}
\Lap{f}(s)\equiv \mathcal{L}[f(t)](s)\! &=\!& \int_0^{\infty} dt e^{-s t} f(t),
\end{eqnarray}
where $s=\gamma - i \omega$ is a complex number ($\gamma$ and $\omega$ are real).
Equation~(\ref{eq:spectrum-time}) can then formally be written as
\begin{equation}
\label{eq:F(omega)}
 F(\omega) = \mathrm{Re} \{\lim_{\gamma\rightarrow 0}\Lap{c}(\gamma - i \omega )\}.
\end{equation}
According to Eq.~(\ref{eq:def:C_abs-time}) one has $\Lap{c}(s)=\sum_{nm}\vec{\mu}_n\cdot\vec{\mu}_m \Lap{\mathcal{C}}_{nm}(s)$.
To obtain the function $\Lap{\mathcal{C}}_{nm}(s)$ one can Laplace-transform the hierarchy Eq.~(\ref{eq:C(t)_HOPS}). 
One finds 
\begin{eqnarray}
-\Psi_n^{(\vec{k})}(t\!=\!\!0)\!
&=\!&\!
-s \, \Lap{\Psi}_n^{(\vec{k})}(s)\nonumber\\
&&-i \sum_m \Big((H_{\rm agg})_{nm} - \delta_{nm}\sum_{j}k_{n j}\omega_{n j}\Big)\Lap{\Psi}_m^{(\vec{k})}(s) \notag\\
&&+\sum_{j}k_{n j}\, p_{n j} \Lap{\Psi}_n^{(\vec{k}-\vec{e}_{n j})}(s) \notag\\
&&-\sum_{j}\Lap{\Psi}_n^{(\vec{k}+\vec{e}_{n j})}(s) 
\label{eq:hops_linear_laplace_nonoise}
\end{eqnarray}
where ${\Psi}_n^{(\vec{k})}(t\!=\!0)$ is the same as for the time-dependent hierarchy.
That means to obtain $\Lap{\mathcal{C}}_{nm}(s)$ one solves the linear system of equations~(\ref{eq:hops_linear_laplace_nonoise}) for a given (complex) frequency $s$ using the initial condition Eq.~(\ref{eq:initial_state}) on the left hand side. 

\subsubsection{Comments on numerical implementation}
As for the time-dependent hierarchy in a numerical implementation a truncation scheme has to be used. 
Since the right hand side of the frequency domain hierarchy (\ref{eq:hops_linear_laplace_nonoise}) has the same structure as that of the time-dependent hierarchy, the number of coupled equations is the same for both hierarchies for a given truncation scheme.
In contrast to the time domain method where the frequency range and frequency resolution are determined by the time-propagation, here one can select arbitrary frequencies at which one wants to evaluate the spectrum. 
We performed calculations for a large range of parameter sets (i.e., different bath correlation functions, and different system Hamiltonians) and found for each parameter set only a very weak dependene of the calculation time on the chosen frequency. 
In this sense the calculation can be trivially parallelized.
In principle the frequency domain hierarchy (\ref{eq:hops_linear_laplace_nonoise}) can be solved by any linear system solver.
However, for large problems (large $N$ and/or large $J$) one has to utilize the fact that the system of equations is sparse.
Therefore iterative methods \cite{Sa03__} are probably the most efficient way to solve  Eq.~(\ref{eq:hops_linear_laplace_nonoise}).

Before concluding this section, some remarks about the limit $\gamma\rightarrow 0$ in Eq.~(\ref{eq:F(omega)}) are in order.
In principle one can set $\gamma=0$ during the evaluation. This is fine for spectra that are broad. 
However, when there are `delta'-like features in the spectrum (as can happen for $\gamma_{nj}=0$ in the bath correlation function) one will most likely miss these features.
Therefore it is recommendable to use a $\gamma$ on the order of (or slightly smaller than) the resolution one wants to have.
This corresponds to the exponential window function in time-domain and leads to a convolution with a Lorentzian in frequency domain.

\section{Conclusion}
One point, relevant for time and frequency formulation, is that the number of coupled equations of hierarchy grows rapidly with the number $J$ of terms used in the representation of the bath-correlation function Eq.~(\ref{eq:def:alpha_decomposition}). 
By decreasing the number of exponentials  the quality of the approximation decreases which also leads to larger deviations from the exact absorption spectrum, in particular for large aggregates.
Therefore, one seeks a fit of the bath-correlation function with $J$ as small as possible, that represents the relevant features of the bath-correlation function (for most molecular aggregates, the spectral density and the bath-correlation function  are also not known with high precision).   
In practice we find it more convenient not to fit the bath-correlation function Eq.~(\ref{eq:def:alpha_decomposition})   directly, but to fit the spectral density $\mathcal{J}(\omega)$ and approximate the coth appearing in  Eq.~(\ref{eq:def:alpha_with_lambda}) by a Pad\'{e} type expansion \cite{RiEi14_094101_}. 
As demonstrated in Ref.~\cite{RiEi14_094101_} this approach results in a decomposition of the bath-correlation function as in Eq.~(\ref{eq:def:alpha_decomposition})  and allows one to handle a large range of relevant spectral densities and finite temperatures by using  quite a small number $J$ of terms.
In  Ref.~\cite{RiEi14_094101_} also several aspects are discussed, regarding the dependence  of the error of the absorption spectrum on the error made when approximating the bath correlation function.
To conclude this discussion, we would like to note, that depending on which features of the absorption spectrum one is (most) interested (e.g., peak-positions, peak intensities, shape of the zero-phonon line, etc.) one can define different measures of the error of the approximated spectrum.

In the present work we have discussed numerical implementation of the hierachy from Ref.~\cite{RiSuMoe15_034115_} which we also presented directly in frequency domain. 
In Ref.~\cite{RiSuMoe15_034115_} the hierarchy is derived starting from an exact {\it stochastic} propagator for the open system.
In contrast, in Ref.~\cite{RoStEi11_034902_} it has been shown that the problem stated in section \ref{sec:Model} with a bath correlation of the form (\ref{eq:def:alpha_decomposition}) can be reformulated such that one enlarges the system by including explicitly harmonic vibrational modes which are coupled to Markovian environments. 
Using a number state representation for these harmonic modes one can derive a hierarchy that has a similar form to the one discussed in the present work.
This hierarchy can be found in Ref.~\cite{RoStEi11_034902_} (see in particular Eq.~(B19) in this reference).
Although the hierarchies have the same structure, they differ by the prefactors appearing in the terms that couple to higher and lower orders.
We have found that large enough orders both hierarchies give the same results.
So far we do not know if there are differences in the speed of convergence with the size of the hierarchy.
Finally, let us note that  the hierarchy of Ref.~\cite{RoStEi11_034902_} can also be formulated in frequency domain.

\end{document}